\documentstyle[12pt]{article}
\begin{document}

\begin{center}
{\large {The $AdS$ particle}} \vskip 2cm
Subir Ghosh\\
\vskip .3cm
Physics and Applied Mathematics Unit,\\
Indian Statistical Institute,\\
203 B. T. Road, Calcutta 700108, India\\
\end{center}
\vskip 2cm
{\bf Abstract:}\\
In this note we have considered a relativistic Nambu-Goto model for
 a particle in $AdS$ metric. With appropriate gauge choice to fix the
 reparameterization invariance, we recover the previously discussed \cite{pal}
 "Exotic Oscillator".  The Snyder algebra and subsequently the $\kappa$-Minkowski
 spacetime are also derived. Lastly we comment on the impossibility of constructing
 a noncommutative spacetime in the context of open string where only  a curved target
 space is introduced.

\vskip 2cm

\newpage

\noindent {\it Introduction}: It is now accepted in the High
Energy Physics community that nonlocality in quantum field theory,
or in a more fundamental way the fuzziness (or Non-Commutativity
(NC)) in space(time), will be an integral part of present-day
theories. Intuitive arguments that are used in avoiding the
paradoxes one faces in trying to localize a spacetime point within
the Planck length \cite{sz} lead to a  lower-bound in spacetime
interval. This feature is also favored in the modifications of the
Heisenberg uncertainty principle that one obtains in  string
scattering results (see for example \cite{sz}). It was first
demonstrated by Snyder \cite{sny} that Lorentz invariance {\it
{and}} discretization requires an NC spacetime.

The NC spacetime has been revived by the seminal work of Seiberg
and Witten \cite{sw} who explicitly demonstrated  the emergence of
NC manifold in certain low energy limit of open strings moving in
the background of a two form gauge field. In this instance, the NC
spacetime is expressed by the Poisson bracket algebra (to be
interpreted as commutators in the quantum analogue),
\begin{equation}
\{x^\mu,x^\nu\}=\theta ^{\mu\nu}, \label{nc0} \end{equation} where
$\theta ^{\mu\nu}$ is a $c$-number constant.  However, quantum
field theories built on this spacetime do not enjoy Poincare
invariance \cite{sub}. On the other hand, this type of pathology
can be avoided if one works with NC spacetime of the Snyder form
\cite{sny} or Lie algebraic form \cite{dfr,ga}. In these examples
the NC is {\it{operatorial}} in nature and thus it does not
jeopardize the Lorentz invariance in relativistic models. The Lie
algebra form of NC spacetime is typically given by,
\begin{equation}
\{x^\mu,x^\nu\}=C^{\mu\nu}_\lambda x^\lambda , \label{nc1}
\end{equation}
 where  the  structure constants $C^{\mu\nu}_\lambda$ are constants.

In the present work, we will encounter both the Snyder \cite{sny}
and Lie algebraic \cite{dfr,ga} forms of NC. In particular, we
will concentrate on
  a restricted class of Lie algebra valued spacetime known
  as $\kappa $-Minkowski spacetime (or $\kappa $-spacetime in short), that is
described by the basic Poisson structure,
\begin{equation}
\{x_i,t\}=kx_i ~~,~~ \{x_i,x_j\}=\{t,t\}=0 .\label{kmin}
\end{equation}
In the above, $x_i$ and $t$ denote the space and time operators
respectively. The present work is in continuation of our recent
paper \cite{pal}.

Some of the important works in $\kappa $-spacetime that discusses,
among other things, construction of a quantum field theory in
$\kappa $-spacetime, are provided in \cite{g1,g2,g3}.
Amelino-Camelia \cite{gam} has pioneered an alternative approach
to quantum gravity - "the doubly special relativity" - in which
{\it two} observer independent parameters, (the velocity of light
and Planck's constant), are present. It has been shown
\cite{kowal} that $\kappa$-spacetime is a realization of the
above. Furthermore, the mapping \cite{kowal} between
$\kappa$-spacetime and Snyder spacetime \cite{sny}, (the first
example of an NC spacetime), shows the inter-relation between
these models and "two-time physics" \cite{bars}, since the Snyder
spacetime can be derived from two-time spaces in a particular
gauge choice \cite{rom}.

In \cite{pal} we have proposed  a physically motivated realization
of the $\kappa$-spacetime in a quantum mechanical model. This is
quite in tune with the connection between the noncommutativity
arising in the Landau problem and that in the open string boundary
with a background field \cite{sz}. It is quite well known that for
the planar, non-relativistic motion of a charged particle in a
magnetic field (in the perpendicular direction), the particle
configuration space becomes effectively noncommutative, if the
dynamics is projected to the lowest Landau level. This is the
celebrated Peirls substitution \cite{pei}. Physically this is
applicable in the limit of strong magnetic field \cite{sz}.
However, it has gained significance in recent times because of its
(qualitative) analogy with the noncommutativity in open string
boundary manifolds ($D$-branes), in the presence of a background
two form gauge field \cite{sw}. Unfortunately, a similar prototype
of a simple physical system, picturizing the $\kappa$-spacetime
was lacking. In our previous paper \cite{pal} we have shed some
light on this area. Specifically, in \cite{pal}, we have put
forward a non-relativistic quantum mechanical model that has an
underlying phase space algebra, isomorphic to the
$\kappa$-Minkowski one (\ref{kmin}). In \cite{pal} we have
provided a Lagrangian of the model. As was mentioned in
\cite{pal}, (this point was noted in \cite{rom} as well), the
action has an uncanny similarity with the structure of the d$S$ or
$AdS$ metric.

Let us put the present work in its proper perspective. The
$\kappa$-spacetime requires the time to be operatorial in nature
since it bears a non-trivial commutation relation with the space
variables as given in (\ref{kmin}). However, our model in
\cite{pal} was non-relativistic with conventional definition of
time. To incorporate the operatorial behavior of time, we had to
convert our model to a generalized one with reparameterization
invariance \cite{ht} and then exploit this symmetry to (gauge)fix
time accordingly so that the $\kappa$-spacetime algebra emerged.
This somewhat roundabout mechanism of \cite{pal}  has led us to
the present work where we extend the non-relativistic particle
model of \cite{pal} to a relativistic, reparameterization
invariant (Nambu-Goto) one. This allows us to fix the form of the
time operator directly in the model. It is interesting to note
that a similar type of time operator as in \cite{pal} reduces the
present model to the one considered in \cite{pal}. We also recover
a generalized form of the Snyder algebra \cite{sny}, first given
in \cite{pal}. But more importantly, now the $AdS$ spacetime comes
in to play directly and hence its connection to the
$\kappa$-spacetime, via the Snyder algebra \cite{sny,pal} and
exotic oscillator \cite{pal} becomes clear. The advantage of
working in a gauge invariant framework is that other convenient
gauge choices, besides the one mentioned above, are indeed
possible.

In an interesting alternative approach, it might be possible to obtain the $\kappa$-spacetime directly from quantum (or noncommutative) $AdS$ spacetime \cite{stein} {\footnote {I thank Professor H. Steinacker for pointing this out.}} One can obtain a broad indication of this connection from the fact that the classical $AdS$-space can be embedded in a higher dimensional space with two-time metric \cite{bars} and the $\kappa$-spacetime is directly related to the latter \cite{kowal,rom}. At a more explicit level, since the $\kappa$-Poincare group can be obtained from the quantum $AdS$ group by contraction \cite{stein}, it is possible the corresponding spaces are related as well.

\vskip .5cm \noindent {\it (Non-relativistic) Mechanical model for
$\kappa$-spacetime}: It will be worthwhile to recapitulate briefly
the model proposed in \cite{pal}. We posited the Lagrangian,
\begin{equation}
L= \frac{m}{2}  \vec{\dot X}^2 - 2mkc\eta (\vec{X}. \vec{\dot X})
+c\eta ^2+2mk^2c^2\eta ^2\vec{X}^2 ,\label{3}
\end{equation}
where $m$ denotes the mass of the non-relativistic particle and
$k$ and $c$ are constant parameters, and as shown below, $\kappa$ and $c$ induce noncommutativity in phase space
related to $\kappa $-spacetime.

In the Hamiltonian constraint analysis, as formulated by Dirac
\cite{dir}, with the
 canonical phase
space,
\begin{equation}
\{X_i,P_j\}=\delta _{ij}~~,~~\{\eta,\pi\}=1 \label{0},
\end{equation}
(where the sets $(X_i,P_j)$ and $(\eta,\pi)$ are decoupled), there
are two  Second Class Constraints (SCC) \cite{dir}{\footnote{In
the Dirac terminology \cite{dir}, First Class Constraints (FCC)
commute with other constraints and generate gauge invariance. We
will come across them in the present work later.}}
\begin{equation}
\chi_1\equiv \pi ~~~,~~~\chi_2\equiv \eta -k(\vec{P}.\vec{X})~.
 \label{1}
\end{equation}
Time independence of $\chi_1$ reproduces $\chi_2$, $(\chi_2
=\{\chi_1,H\})$, with $H$ representing the Hamiltonian.
 SCCs require
the use of Dirac Brackets (DB) \cite{dir} defined by,
\begin{equation}
\{A,B\}_{DB}=\{A,B\}-\{A,\chi _i\}\{\chi _i,\chi _j\}^{-1}\{\chi
_j,B\} \label{db}~,
\end{equation}
such that DB between an SCC and {\it{any}} operator vanishes. Note
that $\{\chi _i,\chi _j\}^{-1}$ indicates inverse of the Poisson
bracket matrix $\{\chi _i,\chi _j\}$. A brief computation
\cite{pal} reveals the following non-canonical Dirac bracket
algebra,
\begin{equation}
\{X_i,\eta \}=kX_i ~~,~~\{P_i,\eta \}=-kP_i~~,~~\{X_i,P_j\}=\delta
_{ij}
 \label{2a}
\end{equation}
Since we will always deal with DBs the subscript DB is dropped.
Clearly $\eta$ behaves as time should in $\kappa$-spacetime, but a
direct identification of $\eta$ with time is obviously not
possible. This was done in \cite{pal} by  extending the model to a
generally covariant. Incidentally, this way of exploiting a
non-standard gauge condition to induce NC coordinates has been
used in \cite{stern} in case of constant spacetime
noncommutativity.

One can eliminate $\eta $ and via an inverse Legendre
transformation, obtain the following Lagrangian:
\begin{equation}
L=P_i\dot X_i-H= \frac{m}{2}[(\dot X_i)^2-(2m\kappa
^2c)\frac{(X_i\dot X_i)^2}{1+(2m\kappa ^2c)X_i^2}]\label{k3}.
\end{equation}
Depending on the sign of $c$, in the context of relativistic point
particle to be demonstrated below, the above expression is
generalized to d$S$ or $AdS$ spacetime. \vskip .5cm \noindent {\it
Relativistic  model for the $AdS$ particle}: The form of the
non-relativistic action in (\ref{k3}) in some sense
 forces up on us its following relativistic counterpart:
\begin{equation}
L=-m[(\dot X.\dot X) -\kappa \frac{(X.\dot X)^2}{1+\kappa (X.X)}
]^{\frac{1}{2}} \equiv -mA  , \label{1a}\end{equation} where
$(X.\dot X)=X^\mu \dot X_\mu $ etc.. Here we have considered a
generic form with a single parameter $\kappa$ and $\dot X^\mu
=\frac{dX^\mu}{d\tau}$. The above Nambu-Goto action clearly has
the built-in $AdS$ metric since the action is
\begin{equation}
{\cal {A}}=\int d\tau \sqrt{g_{\mu\nu}\dot X_\mu \dot X_\nu
}~,~~g_{\mu\nu}=\eta_{\mu\nu}-\frac{\kappa}{1+\kappa X^\lambda
X_\lambda}X^\mu X^\nu . \label{desit}
\end{equation}
The momentum is defined in the usual way,
\begin{equation}
P_\mu \equiv \frac{\delta L}{\delta \dot X^\mu }=-\frac{m}{A}[\dot
X_\mu -\kappa \frac{(X.\dot X)}{1+\kappa X^2 }X_\mu ]. \label{2}
\end{equation}
We directly obtain a modified mass-shell condition
\begin{equation}
(P.P) =m^2-\kappa (P.X)^2, \label{3a}
\end{equation}
 which reduces to the conventional one for $\kappa =0$.  The
 action has a $\tau $-parameterization symmetry that generates a
 zero Hamiltonian:
 \begin{equation}
H=(P.\dot X) -L =0 .\label{4}
\end{equation}
Note that the mass shell constraint in (\ref{3a}) represents the
FCC \cite{dir} We reexpress (\ref{3}) in the form,
\begin{equation}
P_0 =\frac{1}{1+\kappa X_0^2} [\kappa(\vec P.\vec X)X_0 \pm m
(1+\kappa X_0^2)^{\frac{1}{2}}\{1+\frac{\vec
P^2}{m^2}-\frac{\kappa (\vec P.\vec X )^2}{m^2(1+\kappa
X_0^2)}\}^{\frac{1}{2}} ]. \label{5}
\end{equation}
We first demonstrate how the present system reduces to the
non-relativistic model of \cite{pal}. Let us consider the large
$m$ or equivalently the non-relativistic limit,
\begin{equation}
P_0 \approx \frac{1}{1+\kappa X_0^2} [\kappa(\vec P.\vec X)X_0 \pm
m (1+\kappa X_0^2)^{\frac{1}{2}}\{1+\frac{\vec
P^2}{2m^2}-\frac{\kappa (\vec P.\vec X )^2}{2m^2(1+\kappa
X_0^2)}\} ] .\label{6}
\end{equation}
 Keeping terms up to $O(\kappa )$ we rewrite $P_0$ in the following suggestive way,
\begin{equation}
P_0\approx m +\frac{\vec P^2}{2m}-\frac{\kappa}{2m}(\vec P.\vec
X)^2 +\kappa X_0[(\vec P.\vec X)-\frac{m}{2}X_0(1+ \frac{\vec
P^2}{2m^2})].\label{7}
\end{equation}
 Thus, modulo the last term, we have obtained the expression for the Hamiltonian derived in
 \cite{pal}. We can now exploit the reparameterization
 symmetry  to introduce the gauge condition,
\begin{equation}
X_0=\frac{2(\vec P.\vec X)}{m}(1+ \frac{\vec P^2}{2m^2})^{-1} .
\label{8}
\end{equation}
 Clearly the gauge fixed Hamiltonian reduces to that of
 \cite{pal}.

 However, the gauge constraint has rendered the FCC system to an SCC one with the  SCC pair,
\begin{equation}
\phi _1\equiv P_0 -\frac{\vec P^2}{2m}+\frac{\kappa}{2m}(\vec
P.\vec X)^2 + O(\frac{1}{m^3}) ~,~~\phi _2\equiv X_0-\frac{2(\vec
P.\vec X)}{m}+O(\frac{1}{m^3}) . \label{10}
\end{equation}
They   satisfy a non-zero Poisson bracket:
\begin{equation}
\{\phi _1,\phi _2\}=(1- \frac{2\vec P^2}{m^2}) \equiv \alpha
.\label{11}
\end{equation}
The canonical phase space with $\{P_\mu ,X^\nu
\}=\eta^{\nu}_{\mu}$ gets modified to the Dirac brackets,
$$
\{X_i,X_j\}=\frac{2}{\alpha m^2}(X_iP_j-X_jP_i)~~,~~
\{P_i,P_j\}=0$$
\begin{equation}
\{X_i,P_j\}=\delta _{ij}+\frac {2}{\alpha m^2}P_iP_j .\label{12}
\end{equation}
The Dirac brackets with time operator $X^0$ turn out to be,
\begin{equation}
\{X_i,X_0\}=-\frac {2X_i}{\alpha m}~~,~~~\{P_i,X_0\}=\frac
{2P_i}{\alpha m} .\label{time}
\end{equation}

Time evolution is given by the Heisenberg    equations of motion,
\begin{equation}
\dot X_i=\{X_i,P_0\}=\frac {1}{m}(P_i-\kappa(\vec P.\vec
X)X_i)~,~~ \dot P_i=\{P_i,P_0\}=\frac {\kappa}{m\alpha}(\vec
P.\vec X)P_i . \label{14}
\end{equation}
 A further iteration reveals the dynamics:
\begin{equation}
\ddot X_i=-\frac {2\kappa}{m} P_0X_i .\label{15}
\end{equation}
The other equation for $\ddot P_i$  is given below,
\begin{equation}
\ddot P_i=\frac {2\kappa}{m}(\frac{\vec
P^2}{2m}+\frac{\kappa}{2m}(\vec P.\vec X)^2)P_i . \label{16 }
\end{equation}
Thus we have recovered the "Exotic Oscillator" dynamics of
\cite{pal}. A redefinition of the variables, as given in
\cite{pal}, will lead to the $\kappa$-spacetime. The generalized
form of the Snyder algebra, first given \cite{pal}, is also
recovered here in (\ref{12}). Notice that in the approximations
that we have considered, the NC algebra is $\kappa$-independent
but $\kappa$ appears in the dynamics because otherwise we will
have a free particle system.

As we are interested in the $\kappa$-spacetime, quite obviously
the choice of time $(X_0)$ that is obtained from  the form of
gauge fixing is not canonical. Hence it might be interesting to
compare the dynamics with this choice of time and the conventional
($c$-number parameter) one  $X_0=\tau $ by considering an
alternative choice of gauge gauge $\phi_2\approx X_0-\tau $. In
this case, the SCC system is,
$$
\phi_1\approx P_0-[ \frac{\vec P^2}{2m}-\frac{\kappa}{2m}(\vec
P.\vec X)^2 +\kappa \tau[(\vec P.\vec X)-\frac{m}{2}\tau(1+
\frac{\vec P^2}{2m^2})]],$$
\begin{equation}
\phi_2\approx X_0-\tau ,
 \label{17}
\end{equation}
where $\phi_2$ has been used in $\phi _1$. Since now the
Hamiltonian $P_0$ depends explicitly on time $\tau$, one has to
consider the generalized form of Heisenberg equation for a generic
operator $A$,
\begin{equation}
\frac{dA}{d\tau}=\frac{\partial A}{\partial \tau}+\{A,P_0\}.
\label{hei}
\end{equation}
It is clear that the canonical structure  $(\{X_i,P_j\}=\delta_{ij})$
of phase space is not altered by this gauge choice. Thus in case
of conventional time, the dynamics is governed by,
\begin{equation}
\ddot X_i= -\frac{2\kappa}{m}P_0X_i +\kappa(X_i -\frac{\tau
P_i}{m}). \label{18}
\end{equation}
We find that the basic characteristics of the dynamics of the
"Exotic Oscillator" obtained in (\ref{15}) remains intact, since
vanishing of the last term defines the constant non-relativistic
momentum. Perhaps this feature is not so surprising if we recall
that in \cite{pal} the "Exotic Oscillator" dynamics was reproduced
in conventional time with canonical phase space brackets. \vskip
.5cm \noindent {\it{Open String in curved background}}: The next
step in generalization aught to be the de Sitter string that is
string moving in a de Sitter background. However, instead of
considering de Sitter metric in particular, we will consider a
generic form of $X^\mu $-dependent metric $G_{\mu\nu}(X)$. In a
previous work \cite{bcg} we have shown how the boundary conditions
affect the Poisson bracket structures, considering the specific
case of spacetime noncommutativity arising from the non-trivial
boundary conditions occurring in the interacting system of open
string and two-form background gauge field. As a concrete example, in
\cite{bcg} we have shown the noncommutativity
 appearing in the open string boundary manifolds ($D$-branes) in
 the presence of a two-form background field can be rigorously
 obtained once the boundary conditions are properly taken in to
 account. Here we will show that a {\it curved} metric indeed modifies the
 boundary conditions but {\it it does not induce noncommutativity}.

 The canonical phase space algebra
 $$\{X^\mu (\sigma ) ,X^\nu (\sigma ') \}=  \{\Pi_\mu (\sigma),\Pi_\nu (\sigma ')\}=  0~,~~\{\Pi_{\mu}(\sigma) ,X^{\nu}(\sigma ')\}=
 g_{\mu}^{\nu}\delta (\sigma -\sigma '),$$
 is incompatible with the boundary conditions that one obtains for free open
 strings at the boundary and a modified form of $\delta $-function
 $(\Delta (\sigma -\sigma '))$ appears, whose $\sigma$-derivative vanished
 at the string boundary \cite{hr}. On the other hand, for open strings moving
 in the presence of  two-form background field, the modified boundary conditions
 require a non-vanishing $\{X_\mu (\sigma ) ,X_\nu (\sigma ') \}$, indicating
 noncommutativity \cite{bcg}. This point is explained at the end.

The Polyakov action for  the motion of an open string in a curved
background $G_{\mu\nu}(X)$ is,
\begin{equation}
{\cal S}=-\frac{1}{2}\int d\sigma d\tau
\sqrt{-g}g^{ab}\partial_aX^\mu\partial_bX^\nu G_{\mu\nu} ,
\label{20}
\end{equation}
 where $G_{\mu\nu}$ is the curved target space metric and $g_{ab}$ is the
 induced metric. The momentum is defined below:
\begin{equation}
\Pi_\mu =\frac{\delta{\cal S}}{\delta
\partial_0X^\mu}=-\sqrt{-g}G_{\mu\nu}g^{0a}\partial_a X^\nu =-\sqrt{-g}G_{\mu\nu}\partial^0 X^\nu
\label{21}
\end{equation}
Variation of the induced metric $g^{ab}$ determines the
energy-momentum tensor
\begin{equation}
T_{ab} =(-\partial_a X^\mu
\partial_b X^\nu+\frac{1}{2}g_{ab}g^{cd}\partial _cX^\mu \partial_d X^\nu
)G_{\mu\nu}. \label{22}
\end{equation}
Vanishing of the above,
\begin{equation}
T_{ab} =0 \label{22a}
\end{equation}
provides the constraints of the theory that confirms
reparameterization invariance. From the Hamiltonian constraint
analysis point of view, the following combinations of constraints
are useful:
$$
T_{11}=\frac{1}{2}(g\partial^0 X^\mu \partial^0 X^\nu -\partial_1
X^\mu \partial_1 X^\nu )G_{\mu\nu}, $$
\begin{equation}
 \sqrt{-g} {T^0}_1=-\sqrt{-g}\partial^0 X^\mu\partial_1 X^\nu
G_{\mu\nu} .\label{24}
\end{equation}
We can  reexpress the constraints in terms of phase space
variables,
$$
T_{11}\equiv \chi _1 =-\frac{1}{2}(\Pi_\mu \Pi_\nu
G^{\mu\nu}+\partial_1 X^\mu
\partial_1 X^\nu G_{\mu\nu}),$$
\begin{equation}
\sqrt{-g} {T^0}_1\equiv \chi _2=\Pi _\mu \partial_1 X^\mu .
\label{24a}
\end{equation}
The constraints $\chi _1$ and $\chi _2$ are FCC \cite{dir} and
satisfy the normal  diffeomorphism algebra:
$$\{\psi_1(\sigma) ,\psi_1(\sigma ')\}=4(\psi_2(\sigma) +\psi_2(\sigma '))\partial_{\sigma}\delta (\sigma -\sigma '),$$
$$
\{\psi_2(\sigma) ,\psi_1(\sigma ')\}=(\psi_1(\sigma)
+\psi_1(\sigma '))\partial_{\sigma}\delta (\sigma -\sigma '),$$
\begin{equation}
 \{\psi_2(\sigma) ,\psi_2(\sigma ')\}=(\psi_2(\sigma) +\psi_2(\sigma '))\partial_{\sigma}\delta (\sigma -\sigma ').
 \label{diff}
\end{equation}
 Let us now return to the Lagrangian framework. The equation of motion and boundary condition arising from the action (\ref{20}) are respectively,
\begin{equation}
\partial_b [\sqrt{-g}g^{ab}\partial_a X^\mu G_{\mu\nu}]-\frac{1}{2}\sqrt{-g}g^{ab}\partial_a
X^\mu\partial_b X^\lambda\frac{\delta G_{\mu\lambda}}{\delta
X^\nu} =0 ,\label{25}
\end{equation}
\begin{equation}
\sqrt{-g}g^{1a}(\partial _aX^{\mu})G_{\mu\nu}\mid _{\sigma
=0,\pi}=0, \label{25a}
\end{equation}
where $\sigma =0,\pi $ are the string extremities. The boundary
condition, expressed in terms of phase space variables, becomes,
\begin{equation}
\partial _1X^{\mu}+\sqrt{-g}g^{10}\Pi ^{\mu}\mid _{\sigma =0,\pi}=0,
\label{25b}
\end{equation}
where some unimportant factors have been dropped. Notice that the
diffeomorphism algebra (\ref{diff}) ensures  that we can choose
a gauge, in particular the conformal gauge, in which case
$g^{10}=0$, and we are left with $\partial _1X^{\mu}\mid _{\sigma
=0,\pi}=0$ as the boundary condition. This boundary condition is compatible with commutative spacetime. Comparing with our earlier
work \cite{bcg} we establish that spacetime noncommutativity is
not induced by only considering a curved spacetime.

Let us briefly elaborate on the last comment regarding
  \cite{bcg} and its connection to the present conclusion. The
  importance of obtaining the purported noncommutativity from
  different (in particular Hamiltonian) formalisms was stressed in
  the original work of Seiberg and Witten \cite{sw}, since the
  concept of noncommutative spacetime was quite alien to the
  physics community. The first works in this connection
  \cite{chu}, tried to establish that the noncommutative spacetime
  algebra  should be interpreted as Dirac brackets \cite{dir}
  provided one treats {\it{the boundary conditions as Second Class
  Constraints}} \cite{dir}. However, these works \cite{chu} contained
  various assumptions and computational steps that were ambiguous
  from the perspective of conventional Dirac analysis \cite{dir}, \cite{hr} of constrained systems.
  Subsequently it was realized \cite{bcg}, \cite{lor} that the problem lies at the basic
  premises of \cite{chu}: The boundary conditions are {\it{not}} to be treated as  (field theoretic)
  constraints since the former apply {\it{only}} at the boundaries
  whereas the latter are valid for the {\it{whole}} region of phase
  space. This led us to our analysis \cite{bcg} where we generalized
  the earlier works \cite{hr}. It was demonstrated in \cite{hr} that for the case of open strings,  basic phase space Poisson brackets are to be modified in order
  to be consistent with boundary conditions. In \cite{bcg} we
  explicitly showed that the boundary conditions for the
  interacting system of open string in an external two-form gauge
  field are consistent only with a noncommutative spacetime
  algebra. The counter intuitive idea of interpreting boundary
  conditions as constraints as in \cite{chu} need not be
  introduced at all. This explains our conclusion that in the present case that the boundary conditions do not require a noncommutative spacetime.

\end{document}